\documentclass[linenumbers,12pt]{aastex631}
\usepackage{multirow}
\usepackage{color}
\usepackage{amsmath,bm}
\usepackage[percent]{overpic}
\usepackage{rotating}
\usepackage{subfigure}

\definecolor{DarkGreen}{rgb}{0.0,0.4,0.0}  
\newcommand{\B}[1]{{\color{blue}{#1}}}
\newcommand{\R}[1]{{\color{red}{#1}}}
\newcommand{\G}[1]{{\color{DarkGreen}{#1}}}
\newcommand{\M}[1]{{\color{magenta}{#1}}}
\newcommand{\C}[1]{{\color{cyan}{#1}}}
\usepackage[normalem]{ulem}
\usepackage{soul}
\usepackage{cancel}

\newcommand{\erfc}{\operatorname{erfc}}
\newcommand{\e}{\mathrm{e}}
\usepackage[titletoc]{appendix} 
\usepackage{booktabs} 
\usepackage{makecell}

\usepackage{xspace}
\newcommand{\levy}{L\'{e}vy\xspace}
\newcommand{\sar}{SAR 13664/13697\xspace}
\newcommand{\rar}{AR 13679/13711\xspace}

\makeatletter
\AtBeginDocument{%
  \@ifundefined{MakeLineNo}{}{

  }
}
\makeatother

\shorttitle{WTD of solar flares from a global perspective}
\shortauthors{Zhang et al.}

\graphicspath{{./}{figures/}}

\begin{document}

\title{Waiting time distribution of solar flares from a global perspective}

\author[0009-0001-1500-5053]{Yue Zhang}
\affiliation{CAS Key Laboratory of Geospace Environment, Department of Geophysics and Planetary Sciences, University of Science and Technology of China, Hefei 230026,
People’s Republic of China}

\author[0000-0003-4618-4979]{Rui Liu}
\affiliation{CAS Key Laboratory of Geospace Environment, Department of Geophysics and Planetary Sciences, University of Science and Technology of China, Hefei 230026,
People’s Republic of China}
\affiliation{Mengcheng National Geophysical Observatory, University of Science and Technology of China, Hefei 230026, People’s Republic of China}

\author[0009-0006-7051-0438]{Junyan Liu}
\affiliation{CAS Key Laboratory of Geospace Environment, Department of Geophysics and Planetary Sciences, University of Science and Technology of China, Hefei 230026,
People’s Republic of China}

\author[0000-0001-7548-0051]{Dong Wang}
\affiliation{Department of Mathematics and Physics, Anhui Jianzhu University, Hefei, Anhui 230601, People’s Republic of China}

\correspondingauthor{Rui Liu}
\email{rliu@ustc.edu.cn}

\begin{abstract}
The waiting time statistics of solar flares provides clues for the underlying physical mechanisms. However, flares occurring on the far-side have been missing in the statistics. In the 2024 May and June, the Solar Orbiter (SolO) spacecraft orbiting behind the Sun, together with near-Earth spacecrafts, provides a unique opportunity to study one of the most flare-productive active regions, NOAA \sar, over its lifetime, as well as the flare occurrence over the entire solar globe. Derived from time intervals between flare peak times, the waiting time distribution (WTD) is fitted by exponential, log-normal, power-law, and \levy functions with the maximum likelihood estimation method. The goodness of fit is evaluated by the Kolmogorov-Smirnov test, and the statistical models are discriminated by information criteria. The major statistical results are: the WTD of flares in the \sar leans towards the log-normal function, while that in the `normal' \rar towards the \levy function; the WTD of global flares defies the local Poisson hypothesis, and its overall profile cannot be reasonably fitted by any of the four candidate distributions, but its power-law tail $\Delta t^{-\alpha}$ is steeper ($\alpha>3$) than the theoretical expectations ($\alpha\le3$), due to the decreased number of long waiting times ($>10^4$ s) and the increased number of shorter waiting times when the far-side flares are taken into account. These results highlight the importance of studying the flare WTD from a global perspective, and suggest that the long-range magnetic connections in the corona may play a role in the flare occurrences.

\end{abstract}



\section{Introduction} \label{S-intro}
Solar active regions, where photospheric magnetic fields are highly concentrated in sunspots and sunspot groups, are the major energy reservoir for coronal eruptions, such as flares, filament eruptions, and coronal mass ejections (CMEs). In particular, super active regions (SARs) are recognized as large, complex active regions containing large sunspots and producing multiple major flares \citep{Bai1987}. A few key parameters have been introduced to identify SARs, including the maximum area of sunspot group, soft X-ray flare index, 10.7 cm radio peak flux and proton flux, geomagnetic Ap index, and the variation in the total solar irradiance \citep{tian2002, chen2011}. In magnetograms, SARs typically exhibit exceptionally large areas, complex magnetic structures (e.g., $\delta$-spots), and strong magnetic shear along polarity inversion lines \citep{tian2002, Chen&Wang2012, Dhakal&Zhang2024, jaswalDeconstructingPropertiesSolar2025}, and often emerge at certain longitudinal bands on the solar surface \cite[]{Bai1988,tian2002}. Accounting for less than 0.5\% of ARs in a solar cycle, SARs produce more than 40\% of all major flares \cite[]{Bai1987,Bai1988,chen2011,Le2021}. 

On the other hand, it is suggested that the solar corona as a whole is in a state of self-organized criticality \citep[SOC;][]{luAvalanchesDistributionSolar1991a}, just like avalanches in a pile of sands or rice \citep{bakSelforganizedCriticalityExplanation1987a,Frette1996}. When grains keep dropping onto the pile, eventually each new grain can potentially induce an avalanche on any of the relevant size scales after the pile's slope reaches a `critical' number. The adding of grains to the pile is reminiscent of the continual shearing, twisting, and braiding of the coronal magnetic field lines as driven by convective flows in the photosphere, leading to the gradual accumulation of magnetic free energy in the corona and sudden, impulsive energy releases \citep{Priest2014book}. The scale-invariant size distributions of these flaring phenomena lend pivotal support to the SOC theory \citep{Aschwanden2011socbook}.
 
Among the size distributions, waiting time distributions (WTDs) have raised a lot of interest in their revealing the occurrence patterns of many explosive astrophysical phenomena including flares and CMEs \citep[e.g.,][]{Aschwanden2011socbook,kychenthalAlternativeWaitingTime2023,zhangSizeDistributionsArcsecondscale2025}. A comprehensive review of the WTDs of solar flares, CMEs, and radio bursts in the literature concluded that most of them possess a power-law tail $\Delta t^{-\alpha}$ with $\alpha$ in the range 1.5--3.2 \citep{Aschwanden&DudokDeWit2021}. More recently, the power-law indices of WTDs for GOES soft X-ray (SXR) flares in each of the last four solar cycles from 1975 to 2017 are given by \cite{biasiottiStatisticalAnalysisSolar2025}, and they fall in the similar range of 2--3. The power-law tail has implications for SOC \citep{Aschwanden2014}, non-stationary Poisson process \citep{wheatlandOriginSolarFlare2000, Wheatland2003, Aschwanden&McTiernan2010,Aschwanden2021}, or MHD turbulence \citep{boffettaPowerLawsSolar1999a, grigoliniDiffusionEntropyWaiting2002a}. Particularly, it has been controversial as to whether the flare/CME occurrence is independent of, or possesses some memory from, the past \citep{lepretiSolarFlareWaiting2001a,telloniSTOCHASTICITYPERSISTENCESOLAR2013,liWaitingTimeDistributions2018,leiSolarFlaresOriginating2020a,Aschwanden&Johnson2021}. \cite{leiSolarFlaresOriginating2020a} found that the WTDs of flares in the super AR 12673 and in an MHD simulated AR can be well modeled by a continuous, memory-dependent process such as the Stable distribution and the \levy function, but deviate significantly from the discrete, non-stationary Poisson process, suggesting the presence of memory in flare occurrences in individual ARs. It is hence speculated that two mechanisms might dominate at different scales, i.e., the global-wide SOC that is responsible for producing relatively large flares, and the local MHD turbulence for producing a series of flares in an individual AR. 

However, regular observations are limited to the solar disk facing the Earth, which poses a major obstacle to fully understanding WTDs for flares over the entire globe or within an individual AR over its lifetime. Besides the SOC concept that takes the entire corona as a global system, solar eruptions often exhibit long-range correlations \cite[]{schrijverLongrangeMagneticCouplings2011}, coupling through magnetic separatrices or quasi-separatrix layers \citep{Titov2012}. In this letter, taking advantage of two complementary perspectives provided by spacecrafts orbiting the Earth and the Spectrometer Telescope for Imaging X-rays \cite[STIX;][]{Krucker2020} on board the Solar Orbiter \cite[SolO;][]{Mueller2020} spacecraft on the opposite side of the Sun during the lifetime of the \sar spanning over two months, we obtain both the WTD of flares within \sar and of all flares over the entire solar globe. AR 13664 is one of the largest and most active regions in recent decades \citep[e.g., ][]{li2024,jarolim2024, kondrashova2024, hayakawa2025,jaswalDeconstructingPropertiesSolar2025,jing2025}, which makes it a highly valuable observational target for studying the WTD. The rest of the paper is organized as follows: \S\ref{S-method} details the fitting method and the goodness of fit tests. \S\ref{S-results} gives the statistical results. \S\ref{S-discussion} provides concluding remarks.

\section{Methods} \label{S-method}
We define flare waiting times as the peak-to-peak time difference between successive flares \cite[see][for alternative definitions]{kychenthalAlternativeWaitingTime2023}. The resultant WTD is fitted by four different probability density distributions (PDFs) $P(x)$, namely, 
\begin{equation} \label{eq:distributions}
    \begin{split}
    \text{Exponential:\ } &  P_{e}(x) = C\e^{-\lambda x} = \lambda\e^{-\lambda(x-x_{\min})}; \\
    \text{Log-normal:\ }  & P_{l}(x) = C\frac{1}{x}\e{^{-\frac{(\ln{x}-\mu)^2}{2\sigma^{2}}}} = \frac{\sqrt{2/(\pi\sigma^2)}}{\erfc\left(\frac{\ln{x_{\min}-\mu}}{\sqrt2\sigma}\right)} \frac{1}{x\,} \e{^{-\frac{(\ln{x}-\mu)^2}{2\sigma^{2}}}}; \\ 
    \text{Power-law:\ }  & P_{p}(x) = Cx^{-\alpha} = \frac{\alpha-1}{x_{\min}}\left(\frac{x}{x_{\min}}\right)^{-\alpha};\\
    \text{Modified power-law:\ }  &  P_{mp}(x) = C\left(1+\frac{x}{x_\text{tail}}\right)^{-\alpha} = \frac{\alpha-1}{x_\text{tail}}\left(1+\frac{x_{\min}}{x_\text{tail}}\right)^{\alpha-1}\left(1+\frac{x}{x_\text{tail}}\right)^{-\alpha} ;\\
    \text{\levy:\ }  &  P_{v}(x) = C(x-\mu)^{-3/2}e^{-\frac{\gamma/2}{x-\mu}} = \frac{\sqrt{\gamma/2\pi}}{\operatorname{erf}\left(\sqrt{\frac{\gamma/2}{x_{\min}-\mu}}\right)} (x-\mu)^{-3/2} e^{-\frac{\gamma/2}{x-\mu}}.
    \end{split}
\end{equation}
All the above PDFs are normalized for the continuous random variable $x$ over the range $x\ge x_{\min}$ \cite[]{clausetPowerLawDistributionsEmpirical2009a}, with $x_{\min}$ being the specified lower cutoff. $C$ denotes the normalization constant determined by the condition $\int_{x_{\min}}^{\infty} P(x)\,dx = 1$, i.e., $C=[1-F(x_{\min})]^{-1}$, where $F(x)$ is the corresponding cumulative distribution function. $\lambda$ is the characteristic occurrence rate of events in the exponential distribution. $\mu$ and $\sigma$ represent the mean and standard deviation of $\ln x$, respectively, in the log-normal distribution. We use the conventional power-law distribution $P_p$ to fit the tail of a WTD, and a modified version $P_{mp}$ to fit the overall profile. By introducing a characteristic scale parameter $x_\text{tail}$, $P_{mp}$ is intended to account for both the flatten portion of the WTD for small waiting times and the power-law tail for large waiting times. To fit the WTD tail with $P_p$, $x_{\min}$ is given by minimizing the Kolmogorov-Smirnov (KS) distance \cite[]{alstottPowerlawPythonPackage2014}. This is then adopted for tail fittings with other three distributions, and also used as $x_\text{tail}$ for $P_{mp}$. The two-parameter \levy distribution, which is defined for $x \ge x_{\min} > \mu$ with $\gamma$ controlling the width of the distribution and $\mu$ being the location parameter, is a special case of the more general, four-parameter Stable distribution \citep{Nolan2020}. 

We use the maximum likelihood estimation (MLE) method (\texttt{powerlaw.py} in Python) to derive the best-fit parameters directly from the raw data, independent of the binning scheme. The MLE method is known for its robustness, accuracy, and unbiasedness, compared with the least-squares fitting. Adopting a significance level 0.1, we assess the goodness of fit with the KS test: if the resultant $p$-value is smaller 0.1, we reject the fitting distribution. Two competing distributions are discriminated by the log-likelihood ratio \cite[LRT; ][Appendix C]{clausetPowerLawDistributionsEmpirical2009a}. The sign of the ratio $R$ indicates which model is favored, while the statistical significance of this preference is assessed by the $p$-value. For further discrimination, we adopted various information criteria including the Akaike information criterion \cite[AIC; ][]{akaikeNewLookStatistical1974}, where lower AIC values indicate a better balance between goodness of fit and model complexity; the corrected Akaike information criterion \cite[AICc; ][]{sugiuraFurtherAnalysisData1978}, which accounts for finite-sample effects; and the Bayesian information criterion \cite[BIC; ][]{schwarzEstimatingDimensionModel1978}, which imposes a stronger penalty on model complexity.

\section{Results} \label{S-results}

\subsection{Overview of observations} \label{subsec:overview}

\begin{figure} 
\centerline{\hspace*{0.015\textwidth}
\hspace*{-0.03\textwidth}
\includegraphics[width=0.85\textwidth,clip=]{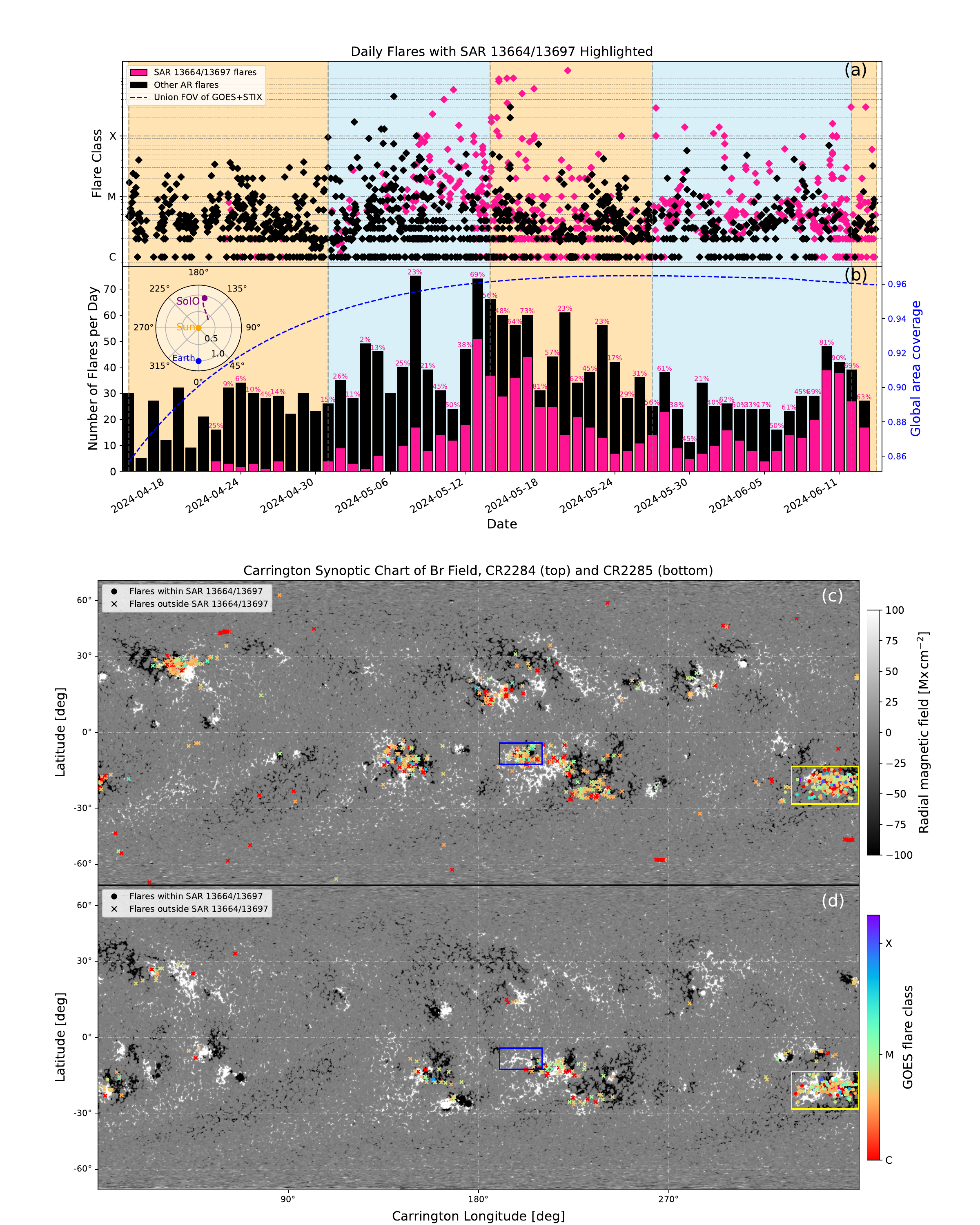}}
\caption{Overview of global flares occurring during the \sar lifetime. Panels (a) and (b) shows the flare magnitude and count number of flares on a daily basis with flares within (outside) \sar  in pink (black) colors. The light-blue and light-orange shades indicate the time interval during which \sar was on the near-side (observed by GOES) and far-side of the Sun (observed by STIX), respectively. The blue dashed curve plots the percentage of the global area covered by the joint perspective of GOES and SolO. The fraction number on the top of bars in (b) gives the daily percentage of \sar flares over all the flares. The inset illustrates in the ecliptic plane the relative positions of the Sun (orange), Earth (blue) and SolO (purple) as well as its trajectory during the \sar lifetime. Panels (c) and (d) show the spatial distribution of flares occurred during the Carrington rotation 2284 (2024 May 6--June 2) and 2285 (June 2--29), respectively. The `o' (`x') symbol marks flares within (outside) \sar. The flare magnitude is color-coded, from strong (purple) to weak (red). The yellow (blue) box marks the range of \sar (\rar).}  
\label{F-C2284+2285}  
\end{figure}

During the period from mid April to mid June 2024, the SolO spacecraft, together with those orbiting the Earth, the Geostationary Operational Environmental Satellite (GOES) and the Solar Dynamic Observatory (SDO), provides an unprecedented opportunity to study the Sun, especially the SAR, from a global perspective (inset of Figure~\ref{F-C2284+2285}b). The far side of the Sun is covered by SolO, and the global area coverage by the joint perspective of Earth and SolO increases from about 85\% to over 95\% during this period (blue curve in Figure~\ref{F-C2284+2285}b). 

To build a global flare catalog, we consulted the flare list from the LMSAL SolarSoft archive\footnote{\url{https://www.lmsal.com/solarsoft/latest_events_archive.html}} and the SolO/STIX flare catalog from the GitHub repository\footnote{\url{https://github.com/hayesla/stix_flarelist_science}}. The flare class in the STIX catalog is given by an empirical correlation between the STIX 4–10 keV flare-peak count rates and the GOES 1--8~{\AA} flare-peak fluxes \citep{xiaoDataCenterSpectrometer2023}. We collected all the GOES and STIX flares that are of C-class and above and occur between 2024 April 15 and June 13. Smaller flares (A- and B-class) are not included in the statistics because they are difficult to be located and are often overwhelmed by the elevated background during active periods. After June 13, the SAR became increasingly diffuse; meanwhile a new AR emerged in close proximity, making it difficult to identify the host AR for quite a few flares occurring in between the two regions. With the IDL GOES software, we pinpointed the GOES flare peak times to the precision of seconds, to match the precision of time stamps in the STIX flare catalog. We corrected the STIX flare peak times to account for the light traveling time from SolO to 1 AU. The flares collected from the two instruments were then chronologically ordered by their peak times. Further, we identified 41 flares that are observed by both instruments and re-assigned them to one of the instruments that records a higher flare magnitude to eliminate repetitive entries. In total, 2,044 C-class-and-above flares occurred over the entire solar globe. 

We then built the flare catalog for the SAR. It first emerged on the solar far side on 2024 April 15, then appeared on the east limb of the near side on 2024 May 1, when it was assigned the AR number 13664. It subsequently rotated over the west limb to the far side on 14 May. When it reappeared on the east limb, it was assigned a new AR number 13697 (hereafter referred to as \sar). Finally it rotated over the west limb to the far side on 12 June 2024, after which the active region further decayed and became a diffused, flare-quiet region. As the STIX flare list does not give the host active region for individual flares, we manually delineated a box spanning from 328 to 360 degrees in Carrington longitude and from -28 to -13 degrees in latitude to enclose the target AR in the Carrington maps covering the two-month period (Fig.~\ref{F-C2284+2285} (c \& d)). We then identified the STIX flares whose locations fall within the specified box as \sar flares. As a sanity check, all but 7 of the GOES flares that are assigned to \sar are also located within the box. In total, this extremely productive SAR produced 741 C-class-and-above flares, including 136 M-class and 34 X-class events. In Figure~\ref{F-C2284+2285} (a \& b), one can see that flares within \sar are dominant over most of May 2024, achieving its SAR status. Hence, the statistics of flares within the SAR would make an effect on that of global flares in the same period. We note in passing that the SXR background level may be different on the opposite side of the Sun. For example, during 2024 May 10–15, the minimum GOES 1–8~{\AA} flux reached a C1.3 level, preventing the detection of C1 flares on the near-side, which is known as the ``obscuration'' effect \citep{wheatland2001}, whereas the STIX background was lower enough to allow the observation of C1 flares on the far-side.

As a reference for comparison, we also built the flare catalog for \rar, which co-existed with the \sar during the same period. This `normal' AR emerged on the far side on April 27, and was assigned the NOAA AR number 13679 when it rotated with Sun to the near side on May 13 as a $\beta\gamma$ type AR. When it rotated to the near side again on June 10 and was given the AR number 13711, it had declined to a $\beta$ type. Similarly, we delineated a box (Figure~\ref{F-C2284+2285}(c \& d)) to enclose \rar in the Carrington map to flag all the STIX flares that occurred within this box. In total we recorded 93 flares within \rar.

\subsection{Waiting Time Statistics}
We first constructed WTDs for flares within individual ARs, namely the \sar and the `normal' \rar for comparison, and then WTDs for the global flares, including three subsets, i.e., those of flare magnitude C-class and above, of M-class and above, and of C-class only. We also compared the WTD of global flares with that excluding the flares in the \sar.

For each WTD, we fit the tail and the overall profile of the distribution separately with the candidate distributions in Eq.~\ref{eq:distributions}. For the tail portion (overall profile) the standard (modified) power law is used. The fitting parameters and the KS test and various information criteria are summarized in Tables~\ref{T-tail} for the tail fitting and in Table~\ref{T-overall} for the overall fitting. Typically the information criteria assign smaller values to the more favored model, as emphasized by the boldfaced entries in the tables.

\subsubsection{WTDs of regional flares} \label{subsubsec:regional_wtd}
\begin{figure} 
\centerline{\hspace*{0.015\textwidth}
\hspace*{-0.03\textwidth}
\includegraphics[width=1.0\textwidth,clip=]{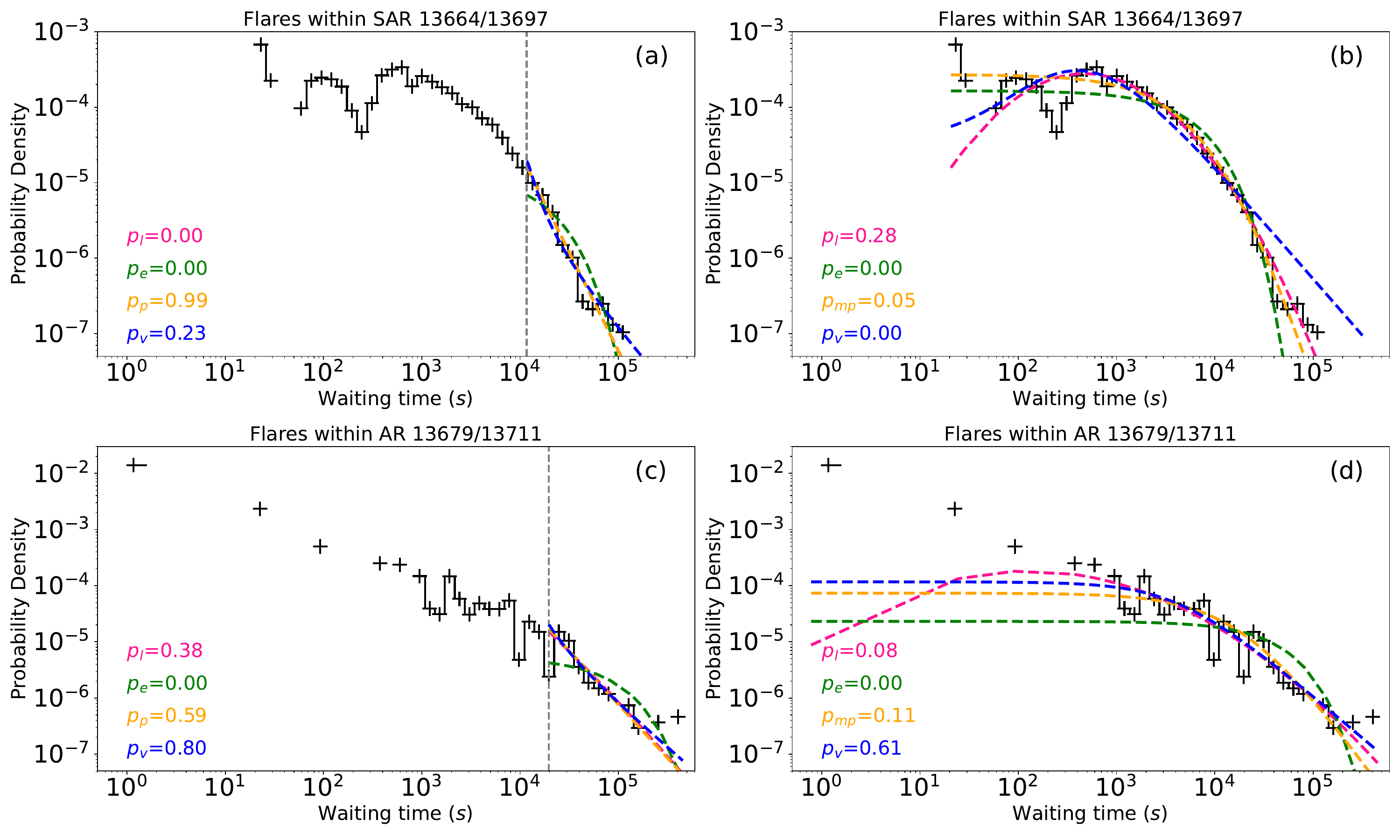}}
\caption{WTDs of flares within individual ARs. Panels (a \& b) show the WTD of flares within \sar; the tail portion is fitted in (a) and the overall profile in (b). The vertical dashed line indicating $x_{\min}$ as determined by minimizing the KS distance for power-law. The $p$-values of the KS tests are shown with subscripts `l', `e', `p'/`mp', and `v', indicating log-normal (pink), exponential (green), power-law (gold), and \levy (blue) functions, respectively. (c \& d) show the WTD of flares within \rar in the same style as (a \& b). }
\label{F-local} 
\end{figure}

The WTDs of flares within individual ARs over their lifetime are shown in Fig.~\ref{F-local}. For flares within the \sar, the tail of the WTD can be well fitted by either power-law or \levy distribution, according to the KS test (Table~\ref{T-tail}). Further tests using the information criteria suggest that the power-law distribution gives a better fitting. The overall profile of the WTD is only well fitted by the log-normal distribution according to the KS test (Table~\ref{T-overall}). 

As a comparison, for flares within the `normal' \rar, the tail of the WTD can be well fitted by log-normal, power-law, or \levy distribution. The KS test excludes only the exponential distribution. Further tests using the information criteria suggests that the \levy distribution provides the best fitting  (Table~\ref{T-tail}). The overall profile of the WTD can be well fitted by either power-law or \levy distribution. The information criteria unanimously vote for the \levy distribution (Table~\ref{T-overall}).

For both the super and normal ARs (Table~\ref{T-tail}), the power-law indices of the WTD tails are below 3. The WTD tail of the \sar is steeper than that of \rar.

\subsubsection{WTDs of global flares} \label{subsubsec:global_wtd}
For all the flares over the entire solar globe (Fig.~\ref{F-global}), the tail portion of the WTD for those occurring outside the \sar can be well fitted by either a log-normal or a power-law distribution. Including flares in the super AR, the WTD tail of C-class flares can be well fitted by either a log-normal or a power-law distribution; that of C-class-and-above flares is preferentially fitted by a power-law distribution; and that of M-class-and-above flares can be well fitted by either a \levy or a power-law distribution. Generally, the tail of global WTDs has a power-law index above 3; the tail portion flattens to an exponent of slightly below 3 if the \sar is artificially excluded (Table~\ref{T-tail}). However, the overall profile of the WTDs of all the above datasets cannot be reasonably fitted by any of the four candidate distributions according to the KS test.

\subsubsection{Local Poissson hypothesis test} \label{subsubsec:H}
We calculated the statistical variable $H$ to test the local Poisson hypothesis \citep{Bi1989,lepretiSolarFlareWaiting2001a,Telloni2014,liWaitingTimeDistributions2018,leiSolarFlaresOriginating2020a}, i.e.,  
\begin{equation}
    H=\frac{2\delta t_i}{2\delta t_i+\delta\tau_i},
\end{equation}
where $\delta t_i = \min \{t_{i+1} - t_i,\ t_i - t_{i-1}\}$ for a series of flares occurring at $t_i$ ($i=1,\ 2,\ 3$, ...), and  
\begin{equation*}
    \delta\tau_i=
    \begin{cases}
     t_{i-1}-t_{i-2} \quad \text{if}\quad \delta t_i = t_i - t_{i-1};\\
     t_{i+2}-t_{i+1} \quad \text{if}\quad \delta t_i = t_{i+1} - t_i.
    \end{cases}     
\end{equation*}
Under the null hypothesis that the flare series is drawn from the Poisson 
distribution, $\delta t_i$ and $\delta\tau_i$ are independent with exponential distributions $P(\delta t_i) = 2\lambda_i\exp(-2\lambda_i\delta t_i)$ and $P(\delta\tau_i) = \lambda_i\exp(-\lambda_i\delta\tau_i)$, where $\lambda_i$ is the non-constant, local event rate. \cite{Bi1989} demonstrated that $H$ is uniformly distributed in $(0,\ 1)$ with a mean value of 0.5. Based on the KS test at the 0.05 significance level, we may reject the local Poisson hypothesis for all the global WTDs, but not for the regional WTDs. 

Further, if an aggregation pattern exists in the time series, $\delta\tau_i$ is typically smaller than $2\delta t_i$, so $H$ is typically larger than 0.5, whereas in a regular pattern, $H$ is typically smaller than 0.5. From Table~\ref{T-overall}, one can see that varying degrees of clustering is found in all WTDs except the one for large flares (M-class and above), and that \sar has the largest mean/median $H$ value among all the WTDs, indicating that the flare occurrence in this super AR is highly clustered. 

\section{Discussion and Conclusion} \label{S-discussion}
In this study, taking advantage of the unprecedented global vision provided by the joint observation of GOES and SolO, we conducted a comprehensive analysis of the WTD of solar flares, including flares within the \sar over its lifetime and those over the rest of the entire solar globe during the same period. 

The waiting time statistics reveals a subtle difference between the \sar, which registers 741 flares of C-class and above, and the `normal' \rar, which registers 93 flares of C-class and above. The WTD of flares within the former AR slightly prefers the power-law to the \levy distribution in the tail (Table~\ref{T-tail}), but overall it can only be reasonably fitted by the log-normal distribution (Table~\ref{T-overall}). In contrast, the WTD of flares within the latter AR prefers the \levy distribution in both the tail and overall fittings. Similarly, the WTD of the super AR 12673, which produces 158 C-class-and-above flares during its disk passage from 2017 August 30 to September 11, can also be fitted by the \levy function \citep{leiSolarFlaresOriginating2020a}. On the one hand, we argue that the waiting time statistics is more robust in the \sar due to the large number of flares, which suggests that the WTD of flares within individual ARs is more likely to be log-normal than \levy. On the other, when the flare frequency increases, it also leads to more and more time-overlapping events, or ``event pileup'' \citep{Aschwanden2014}, so higher $H$ values (\S\ref{subsubsec:H}). In this occasion, the waiting time statistic could be `contaminated' by the distribution of flare durations. With three different definitions of waiting times, i.e., the end-to-start, peak-to-peak, and start-to-start times of consecutive events, \cite{kychenthalAlternativeWaitingTime2023} revisited the classic avalanche model for flares \citep{luAvalanchesDistributionSolar1991a} and found that the distribution of the end-to-start waiting time, which may exclude the influence of avalanche durations if there is no clustering, is exponential, but power-law for the other two definitions. Thus, it requires more careful studies to clarify the WTD of flares within individual ARs.

In regard to the statistics of global flares, the WTDs generally display a heavy tail, which not only deviates from the exponential distribution, but also defies the local Poisson hypothesis. In addition, the WTD tail of M-class flares seem flatter and more extended into longer waiting times than that of C-class flares; the median waiting times with 95\% confidence intervals \citep{biasiottiStatisticalAnalysisSolar2025} for C-, M-, and X-class flares are $0.5\pm0.1$, $2.3\pm0.9$, and $11.2\pm11.6$ hrs, respectively, which is in accordance with the `build-up and release' scenario for coronal eruptions \citep{Hudson2020MNRAS}. However, investigations on flares within the same active region generally found no correlation between waiting time and flare peak-intensity \citep{Wheatland2000,Hudson2020MNRAS,Hudson2020SoPh,biasiottiStatisticalAnalysisSolar2025}. As for the full-disk statistics, \cite{biasiottiStatisticalAnalysisSolar2025} reported that the median waiting times are nearly constant for flares of different SXR classes over Solar Cycles 21--24. \cite{liWaitingTimeDistributions2018} showed that the WTD of weak flares is different from that of strong flares, with the latter being closer to a stochastic process. A caveat to keep in mind, however, is that the detection of weak flares could be significantly affected by the elevated SXR background during active periods, especially by the ``obscuration'' effect of long-duration flares \citep{wheatland2001}. Future investigations will tackle the relationship between waiting time and flare magnitude from the global perspective.

For the overall profile, none of the candidate distributions is able to adequately describe the global WTDs, which might be a manifestation of the intrinsic complexity of flare activities. From a global perspective, the flares originate from multiple host active regions with diverse magnetic configurations and evolutionary paths, and unlike grains in the prototypical SOC models \citep{bakSelforganizedCriticalityExplanation1987a} there may exist long-range magnetic connections between them \citep{schrijverLongrangeMagneticCouplings2011,Titov2012}, therefore leading to the entanglement of multiple complex processes \citep{Uritsky2007}.  

\begin{figure} 
\centerline{\hspace*{0.015\textwidth}
\hspace*{-0.03\textwidth}
\includegraphics[width=1.0\textwidth,clip=]{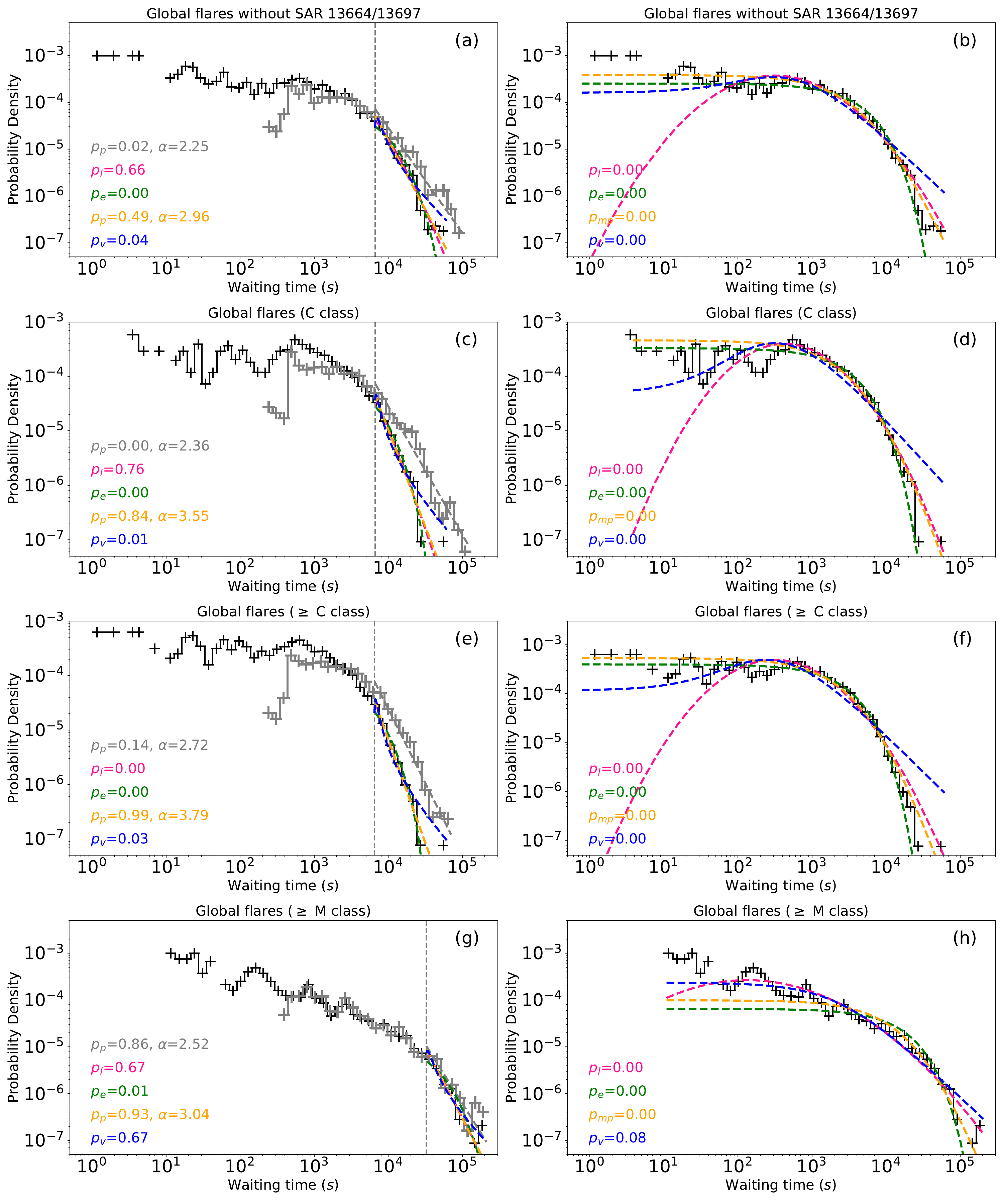}}
\caption{WTDs of global flares. The format is the same as Figure~\ref{F-local}. For the WTD in the same row, left (right) panel shows the fitting of the tail portion (overall profile). In the left column, the corresponding WTD of near-side flares observed by GOES is shown in gray; its tail, starting from the same $x_{\min}$ as that of the global WTD, is fitted by power law, and the resultant $p$-value and power-law index are annotated.} 
\label{F-global}  
\end{figure}

Further, the tail of WTDs for global (regional) flares generally has a power-law index above (below) 3. In contrast, previous studies invoking the SOC phenomena in a global wide system \citep{Aschwanden2014}, or a nonstationary Poisson process \citep{Wheatland2003, Aschwanden&McTiernan2010, Li2014, Aschwanden2021}, or MHD turbulence \citep{boffettaPowerLawsSolar1999a,lepretiSolarFlareWaiting2001a,grigoliniDiffusionEntropyWaiting2002a} typically generate a WTD with a power-law tail $\sim\Delta t^{-\alpha}$ ($\alpha\leq3$). In comparison with the WTD of the near-side flares observed by GOES only (left column in Fig.~\ref{F-global}), the WTD of global flares is extended substantially into the region of short waiting times (below $\sim10^2$~s), enhanced in the region of medium waiting times ($\sim10^3$–-$10^4$~s), and steepened in the region of long waiting times above $10^4$~s. Obviously, when the far-side flares are taken into account, waiting times tend to be split into shorter intervals, which causes the above-mentioned differences between the full-disk and the global WTD. Unlike the ``event pile-up'' during busy periods as observed by a single spacecraft \citep{Aschwanden2014}, here the time overlapping between near-side and far-side events does not lead to the underestimation of event durations. Thus, the difference in the power-law tail between the global WTD and the theoretical expectations calls for a re-examination of the physical mechanism behind the flare occurrence, as long as the solar corona is treated as a whole.

\begin{table}
\centering
\caption{Goodness of fit for WTDs of flares (tail fitting)}

\begin{tabular}{ccccc|cccc}
\toprule
&  \multirow{2}{*}{Model}   &     & \multicolumn{2}{c}{WTD of regional flares}  & \multicolumn{4}{c}{WTD of global flares}   \\
&  &   & \rar & \sar  & w/o SAR &  C-class  & $\geq$\ C1 & $\geq$\ M1 \\\midrule
\multirow{6}{*}{params}  & \multirow{2}{*}{$P_l$} & $\mu$     & 2.88       & -210.4           & 7.25             & 4.11             & -214.54              & 0.17                 \\
                         &                        & $\sigma$  & 3.16       & 11.70            & 1.12             & 1.46             & 8.96                 & 2.34                 \\
                         & $P_e$                  & $\lambda$ & 1.10E-05   & 5.80E-05         & 1.72E-04         & 2.45E-04         & 2.75E-04             & 3.50E-05             \\
                         & $P_p$                  & $\alpha$  & 1.91       & 2.61             & 2.96             & 3.55             & 3.79                 & 3.04                 \\
                         & \multirow{2}{*}{$P_v$} & $\mu$     & 8.87E+03   & 1.05E+04         & 6.27E+03         & 6.36E+03         & 6.31E+03             & 3.03E+04             \\
                         &                        & $\gamma$  & 1.00E-08   & 3.42E+03         & 2.10E+03         & 1.38E+03         & 1.15E+03             & 1.03E+04             \\
                         \midrule
\multirow{4}{*}{KS$^{\rm a}$} & $P_l$             &        & 0.38  & 0.00    & 0.66    & 0.76    & 0.00    & 0.67        \\
                         & $P_e$                  &        & 0.00   & 0.00   & 0.00    & 0.00    & 0.00    & 0.01        \\
                         & $P_p$                  &        & 0.59  & 0.99    & 0.49    & 0.84    & 0.99    & 0.93        \\
                         & $P_v$                  &        & 0.80  & 0.23    & 0.04    & 0.01    & 0.03    & 0.67        \\
                         \midrule
\multirow{6}{*}{LRT$^{\rm b}$}     & $P_l$ vs $P_e$    &        & \textbf{10.21 (0.01)}  & \textbf{21.20 (0.05)}     & \textbf{9.91 (0.05)}      & 11.25 (0.13)     & 14.39 (0.11)         & \textbf{4.33 (0.08)}          \\
                         & $P_l$ vs $P_p$              &        & 0.15 (0.67)   & 0.00 (0.99)      & 2.34 (0.18)      & 0.36 (0.66)      & -0.00 (0.95)         & 0.04 (0.84)          \\
                         & $P_e$ vs $P_p$              &        & \textbf{-10.06 (0.01)} & \textbf{-21.20 (0.05)}    & -7.57 (0.27)     & -10.90 (0.18)    & -14.39 (0.11)        & -4.29 (0.11)         \\
                         & $P_l$ vs $P_v$              &        & -3.48 (0.82)  & 3.01 (0.89)      & 6.16 (0.79)      & 12.16 (0.65)     & 9.91 (0.69)          & -1.77 (0.89)         \\
                         & $P_e$ vs $P_v$              &        & -13.69 (0.29) & -18.19 (0.48)    & -3.75 (0.88)     & 0.91 (0.97)      & -4.48 (0.87)         & -6.09 (0.65)         \\
                         & $P_p$ vs $P_v$              &        & -3.63 (0.81)  & 3.01 (0.89)      & 3.82 (0.87)      & 11.80 (0.66)     & 9.91 (0.69)          & -1.80 (0.89)         \\
                         \midrule
\multirow{4}{*}{AIC}     & $P_l$                  &        & 825.60  & 1831.99          & \textbf{4430.94} & \textbf{3650.74} & 2973.63              & 1053.91              \\
                         & $P_e$                 &        & 844.02  & 1872.39          & 4448.75          & 3671.25          & 3000.42              & 1060.57              \\
                         & $P_p$                   &        & 823.90  & \textbf{1829.99} & 4433.62          & \textbf{3649.46} & \textbf{2971.63}              & \textbf{1051.99}     \\
                         & $P_v$                       &        & \textbf{817.64}  & 1838.01          & 4443.26          & 3675.07          & 2993.46              & \textbf{1050.38}     \\
                         \midrule
\multirow{4}{*}{AICc}    & $P_l$                  &        & 825.99  & 1832.13          & \textbf{4430.99} & \textbf{3650.80} & 2973.71              & 1054.19              \\
                         & $P_e$                 &        & 844.14  & 1872.44          & 4448.77          & 3671.27          & 3000.44              & 1060.65              \\
                         & $P_p$                  &        & 824.02  & \textbf{1830.04} & 4433.64          & \textbf{3649.48} & \textbf{2971.66}              & \textbf{1052.07}     \\
                         & $P_v$                       &        & \textbf{819.03}  & 1838.15          & 4443.31          & 3675.13          & 2993.53              & \textbf{1050.65}     \\
                         \midrule
\multirow{4}{*}{BIC}     & $P_l$                  &        & 828.65  & 1836.92          & \textbf{4437.81} & 3657.31          & 2979.82              & 1057.61              \\
                         & $P_e$                 &        & 845.54  & 1874.86          & 4452.19          & 3674.53          & 3003.51              & 1062.42              \\
                         & $P_p$                   &        & 825.43  & \textbf{1832.45} & \textbf{4437.06} & \textbf{3652.74} & \textbf{2974.73}              & \textbf{1053.84}     \\
                         & $P_v$                       &        & \textbf{821.69}  & 1842.94          & 4450.14          & 3681.63          & 2999.65              & \textbf{1054.08}     \\
\midrule
& Favored   &  & $P_v$  & $P_p$  & $P_l$ or $P_p$  & $P_l$ or $P_p$  &  $P_p$  & $P_l$ or $P_v$    \\
\bottomrule
\end{tabular}
\raggedright\small{
$^{\rm a}$$p$-value of the Kolmogorov-Smirnov test.\\
$^{\rm b}$Log-likelihood ratio between two candidate distributions and the corresponding $p$-value inside the brackets.}
\label{T-tail}  
\end{table}

\begin{table}[]
\centering
\setlength{\tabcolsep}{4pt}
\caption{Goodness of fit for WTDs of flares (overall fitting)}
\begin{tabular}{ccccc|cccc}
\toprule
&  \multirow{2}{*}{Model}   &     & \multicolumn{2}{c}{WTD of regional flares}  & \multicolumn{4}{c}{WTD of global flares}   \\
&  &   & \rar & \sar  & w/o SAR &  C-class  & $\geq$\ C1 & $\geq$\ M1 \\\midrule
                         & $H$ (mean/median)            &          & 0.50/0.52     & 0.52/0.55           & 0.51/0.52             & 0.52/0.53          & 0.51/0.51           & 0.45/0.40                 \\
                         \midrule
\multirow{6}{*}{params}  & \multirow{2}{*}{$P_l$}      & $\mu$    & 9.10     & 7.85           & 7.59             & 7.44          & 7.25           & 8.46                 \\
                         &                             & $\sigma$ & 2.06     & 1.30           & 1.35             & 1.18          & 1.22           & 1.89                 \\
                         & $P_e$                       & $\lambda$& 2.30E-05 & 1.65E-04       & 2.52E-04         & 3.30E-04      & 3.95E-04       & 6.50E-05             \\
                         & $P_{mp}$                    & $\alpha$ & 2.45     & 4.14           & 3.53             & 4.05          & 4.43           & 4.22                 \\
                         & \multirow{2}{*}{$P_v$}      & $\mu$    & -2209.87 & -139.91        & -215.72          & -114.34       & -114.89        & -2388.25             \\
                         &                             & $\gamma$ & 4719.73  & 1610.34        & 1541.26          & 1282.63       & 1062.58         & 1.00E-08             \\
                         \midrule
\multirow{4}{*}{KS}      & $P_l$                       &          & 0.08     & 0.28           & 0.00             & 0.00          & 0.00           & 0.00                 \\
                         & $P_e$                       &          & 0.00     & 0.00           & 0.00             & 0.00          & 0.00           & 0.00                 \\
                         & $P_{mp}$                    &          & 0.11     & 0.05           & 0.00             & 0.00          & 0.00           & 0.00                 \\
                         & $P_v$                       &          & 0.61     & 0.00           & 0.00             & 0.00          & 0.00           & 0.08                 \\
                         \midrule
\multirow{6}{*}{LRT}     & $P_l$ vs $P_e$              &          & \textbf{39.98 (0.00)}     & \textbf{127.86 (0.01)}   & -34.51 (0.15)    & -12.02 (0.63) & \textbf{-71.58 (0.02)}        & \textbf{41.84 (0.02)}         \\
                         & $P_l$ vs $P_{mp}$           &          & -2.74 (0.91)     & 3.02 (0.96)     & -63.15 (0.27)    & -14.42 (0.83) & -75.20 (0.30)        & 4.96 (0.90)          \\
                         & $P_e$ vs $P_{mp}$           &          & \textbf{-42.72 (0.08)}    & \textbf{-124.84 (0.03)}  & -28.64 (0.61)    & -2.40 (0.97)  & -3.62 (0.96)         & -36.88 (0.35)        \\
                         & $P_l$ vs $P_v$              &          & -7.75 (0.74)     & 70.41 (0.19)    & 44.32 (0.42)     & \textbf{169.75 (0.01)} & \textbf{204.26 (0.00)}        & -6.33 (0.87)         \\
                         & $P_e$ vs $P_v$              &          & \textbf{-47.73 (0.05)}    & -57.45 (0.46)  & 78.84 (0.16)    & \textbf{181.77 (0.01)}  & \textbf{275.84 (0.00)}   & -48.17 (0.22)        \\
                         & $P_{mp}$ vs $P_v$           &          & -5.01 (0.18)     & \textbf{67.39 (0.00)}    & \textbf{107.48 (0.00)}    & \textbf{184.17 (0.00)} & \textbf{279.46 (0.00)}        & -11.29 (0.24)        \\
                         \midrule
\multirow{4}{*}{AIC}     & $P_l$                       &          & 2072.21     & \textbf{14121.90}        & 24258.84         & 30840.86      & 36256.44             & 7010.75              \\
                         & $P_e$                       &          & 2150.18     & 14375.62        & 24187.81         & 30814.82      & 36111.27             & 7092.43              \\
                         & $P_{mp}$                    &          & 2064.73     & 14125.94        & 24130.53         & 30810.01      & 36104.03             & 7018.67              \\
                         & $P_v$                       &          & \textbf{2056.72}     & 14262.73        & 24347.48         & 31180.35      & 36664.95             & 6998.09              \\
                         \midrule
\multirow{4}{*}{AICc}    & $P_l$                       &          & 2072.35     & \textbf{14121.91}        & 24258.85         & 30840.87      & 36256.44             & 7010.79              \\
                         & $P_e$                       &          & 2150.22     & 14375.63        & 24187.81         & 30814.82      & 36111.27             & 7092.44              \\
                         & $P_{mp}$                    &          & 2064.78     & 14125.94        & 24130.53         & 30810.01      & 36104.03             & 7018.68              \\
                         & $P_v$                       &          & \textbf{2056.85}     & 14262.74        & 24347.49         & 31180.36      & 36664.96             & 6998.13              \\
                         \midrule
\multirow{4}{*}{BIC}     & $P_l$                       &          & 2077.26     & \textbf{14131.11}        & 24269.18         & 30851.75      & 36267.68             & 7018.37              \\
                         & $P_e$                       &          & 2152.70     & 14380.23        & 24192.98         & 30820.26      & 36116.89             & 7096.24              \\
                         & $P_{mp}$                    &          & 2067.26     & 14130.54        & 24135.70         & 30815.45      & 36109.65             & 7022.48              \\
                         & $P_v$                       &          & \textbf{2061.76}     & 14271.94        & 24357.83         & 31197.24      & 36676.20             & 7005.71              \\
                         \midrule
 & Favored                             &          & $P_v$     & $P_l$      &                  &               &                      &         \\
\bottomrule
\end{tabular}
\label{T-overall} 
\end{table}

\begin{acknowledgments}
This work was supported by the Strategic Priority Program of the Chinese Academy of Sciences (XDB0560102), the National Key R\&D Program of China (2022YFF0503002), and the NSFC (42274204, 42188101, 11925302). 
\end{acknowledgments}


\bibliography{SAR}
\bibliographystyle{aasjournal}




\end{document}